\documentclass[prl,twocolumn,superscriptaddress,showpacs]{revtex4}

\newcommand{\lvec}{\mathbf{l}}

\usepackage{graphicx}

\begin{document}

\title{Trapping of lattice polarons by impurities}

\author{J.P. Hague}
\affiliation{Department
 of Physics, Loughborough University, Loughborough, LE11 3TU, United Kingdom}
\affiliation{Department
 of Physics and Astronomy, The Open University, Milton Keynes, MK7 6AA, United Kingdom}

\author{P.E. Kornilovitch}
\affiliation{Hewlett-Packard Company, 1000 NE Circle Blvd,
Corvallis, Oregon 97330, USA}

\author{A.S. Alexandrov}
\affiliation{Department
 of Physics, Loughborough University, Loughborough, LE11 3TU, United Kingdom}

\begin{abstract}

We consider the effects of single impurities on polarons in
three-dimensions (3D) using a continuous time quantum Monte-Carlo
algorithm. An exact treatment of the phonon degrees of freedom leads
to a very efficient algorithm and we are able to compute the polaron
dynamics on an infinite lattice using an auxiliary weighting
scheme. The magnitude of the impurity potential, the electron-phonon
coupling and the phonon frequency are varied. We determine the
magnitude of the impurity potential required for polaron trapping.
For small electron-phonon coupling the number of phonons increases
dramatically on trapping. The polaron binding diagram is computed,
showing that intermediate-coupling low-phonon-frequency
polarons are localized by exceptionally small impurities.

\end{abstract}

\maketitle

Interest in the role of electron-phonon interactions (EPIs) and
polaron dynamics  has recently gone through a vigorous revival.
Polarons  have been shown to be relevant in high-temperature
superconductors, colossal magnetoresistance oxides, polymer and many
inorganic semiconductors, and electron transport through nanowires
 often depends on vibronic displacements of ions.  The continued
interest in polarons  extends well beyond the physical description of
advanced materials. The field has been a testing ground for
analytical, semi-analytical, and different numerical techniques
\cite{polarons}.

The situation as regards polaron formation and dynamics in real
materials is complicated by an intrinsic disorder. Moreover, the EPI
cannot be considered either weak or strong in many of the materials
described above, so standard approximations based on perturbation
theories break down. In a pioneering paper Economou and coauthors
\cite{economou1993a} studied a one-dimensional (1D) large polaron with
diagonal disorder by using methods from the theory of non-linear
systems. Bronold \emph{et al.}  \cite{bronold2001a} investigated the
dynamics of a single electron in a Holstein model with a
site-diagonal, binary-alloy-type disorder by applying a dynamical
mean-field theory (DMFT) for a Bethe lattice with infinite
coordination number. DMFT was further improved by Bronold and Fehske
\cite{bronold2002}, who described Anderson localization and the
self-trapping phenomena within the same model.

There is a delicate interplay between the self-trapping by EPI and
trapping by doped-induced disorder in the intermediate-coupling
regime, so even a single polaron has to be studied by
 exact methods such as continuous-time quantum Monte-Carlo
(CTQMC)  \cite{kornilovitch1998a, hague2007a} and/or diagrammatic
Monte-Carlo (DMC) \cite{mishchenko} techniques. In particular, the CTQMC
algorithm treats the phonon degrees of freedom exactly, does not
suffer from time discretization errors and is not limited to finite
lattices, thus providing numerically exact solution of the (bi)polaron
problem in any dimensions and on any lattice.

In this paper, we use CTQMC to solve the problem  of the
electron interacting with phonons in the presence of an impurity on
a 3D lattice.  The
  electron hopping is assumed to be between nearest neighbors
  only.  The phonon subsystem is made up of independent
  oscillators with frequency $\omega$, displacement $\xi_{\bf m}$, momentum $\hat{P}_{\bf m} = -i\hbar\partial/\partial \xi_{\bf m}$ and mass $M$ associated with each lattice site. The real space
 Hamiltonian reads,
\begin{eqnarray}
H & = & - t \sum_{\langle \mathbf{nn'} \rangle}
c^{\dagger}_{\mathbf{n'}} c_{\mathbf{n}}+ \sum_{
\mathbf{n}} \Delta_{\mathbf{n}}c^{\dagger}_{\mathbf{n}}
c_{\mathbf{n}} \label{eq:hamiltonian} \\
&&+ 
\sum_{\mathbf{m}} \left(\frac{\hat{P}^{2}_\mathbf{m} }{2M}
+ \frac{M\omega^2\xi^{2}_{\mathbf{m}}}{2}
  \right) -
\sum_{\mathbf{n}\mathbf{m}} f_{\mathbf{m}}(\mathbf{n})
c^{\dagger}_{\mathbf{n}} c_{\mathbf{n}} \xi_{\mathbf{m}}
\: .\nonumber 
\end{eqnarray}
Here $\langle \bf{nn'} \rangle$ denote pairs of nearest neighbors. The
  sites are indexed by ${\bf n}$ or ${\bf m}$ for electrons and ions
  respectively. The spin indices and Hubbard $U$ are omitted since
  there is only one electron.  The strength of the EPI is defined
  through a dimensionless coupling constant, $\lambda = \sum_{\bf m}
  f^2_{\bf m}(0)/2M \omega^2 zt$ which is the ratio of the polaron
  energy when $t = 0$ to the kinetic energy of the free electron $W =
  zt$. In this paper, we discuss Holstein polarons, which have a force
  function, $f_{\bf m}({\bf n})=\kappa\delta_{\bf n,m}$
  \cite{holstein1959a}. A special case of the Hamiltonian given in
  eq. (\ref{eq:hamiltonian}) is the case where the external
  potential, $\Delta_{\bf n} = \Delta \delta_{\bf n,0}$, representing
  an impurity. We will study this in detail in this paper. We note
  that there is a significant difference between this problem, where
  electrons in both host and impurity are coupled to phonons and the
  traditional electron-phonon impurity problem, where only the
  impurity site is coupled to a phonon mode \cite{hewsonprl}. In the
  problem considered here, polarons can exist in both the host and
  impurity.

The CTQMC method employed here has been described in detail with
regard to the single polaron problem in Ref.
\onlinecite{kornilovitch1998a}. Here we give a quick overview of the
extended algorithm for impurities, which is one of the main
developments here.  The effective polaron action that results when the
phonon degrees of freedom have been integrated out analytically is
given by,
\begin{widetext}
\begin{equation}
A[{\bf r}(\tau)]  =  \frac{z\lambda\bar{\omega}}{2}
\int_0^{\bar\beta} \int_0^{\bar\beta} d \bar{\tau} d \bar{\tau}'
e^{-\bar{\omega} \bar\beta/2}\ \cosh(\bar{\omega}(\bar\beta/2-|\bar{\tau}-\bar{\tau}'|))
\frac{\Phi_0[\mathbf{r}(\tau),\mathbf{r}(\tau')]}{\Phi_0(0,0)}
 -\int_0^{\beta}\Delta(\mathbf{r}(\tau))\,d\tau \: ,
 \label{eq:seven}
\end{equation}
\end{widetext}
The contribution of this action to the statistical weight of a path
configuration is $e^{A}$.  
$ \Phi_{0}[{\bf r}(\tau), {\bf r}(\tau')] =
\sum_{\bf m} \bar{f}_{\bf m}[{\bf r}(\tau)] \bar{f}_{{\bf m}} [{\bf
r}(\tau')] $
(here $\bar{\omega}=\omega/t$ and $\bar\beta=t/k_BT$) is the phonon
mediated interaction.

 One of the main complications regarding the simulation of a particle
in a single impurity potential is ensuring that the whole
configuration space is sampled. This can be understood in the
following way. During a binary kink update (discussed below) the ends
of the path may either stay put or move right or left by one site
along one of the nearest neighbor bonds, i.e. the path is essentially
a random walker. We set up the system with a very small attractive
$\Delta\rightarrow 0^{-}$, with the path close to the impurity. In 1D,
the random walker has finite probability to return to the start site
after a finite number of steps. In 2D the walker has vanishing
probability to return, so ergodicity is not guaranteed. In 3D, the
walker will not in general return to its start point {\bf even} after
an infinite time, so if updates only have the properties of a basic
random walker, then the Monte-Carlo procedure will fail, since the
impurity will (on average) never be visited.

In order to ensure that the path includes sufficient sampling of the
impurity, it would be useful to make the path spend more time near the
impurity, weighting the estimators in such a way that the final
measurements are the same. Sampling in this way may be achieved by
introducing an auxiliary weighting to the problem.  A functional of
the path $w[\{C\}]$ in introduced, which represents the probability
that an unbound path is found in a particular configuration, which is
tuned so that the path samples the impurity regularly. $w[\{C\}]$
modifies the measurements as,
\begin{eqnarray}
\langle O \rangle & = & \left\langle \hat{O} \right\rangle_{A_{\rm tot}} = \frac{\sum_{\{
C\}} O[\{C\}] e^{A_{\rm tot}[\{C\}]}}{\sum_{\{
C\}} e^{A_{\rm tot}[\{C\}]}} \\
& = & \frac{\sum_{\{
C\}} O[\{C\}] w[\{C\}]  e^{A_{\rm tot}[\{C\}]}/w[\{C\}] }{\sum_{\{
C\}} w[\{C\}] e^{A_{\rm tot}[\{C\}]}/w[\{C\}] } \\
& = & \frac{\left\langle \hat{O}/w[\{C\}] \right\rangle_{A_{\rm tot},w}}{\left\langle 1/w[\{C\}] \right\rangle_{A_{\rm tot},w}}
\end{eqnarray}
and the update probabilities are modified by the ratio of weights
$w[\{D\}]/w[\{C\}]$ on changing from configuration ${C}$ to ${D}$.
$A_{\rm tot}$ is the total action including kinetic energy terms (for
reasons relating to the continuous time algorithm, the kinetic energy
part of the statistical weight is taken into account through the
update rule, eq. \ref{eqn:update}). Thus the measurement of an
observable with respect to the ensemble defined by the action $A_{\rm
tot}$ can be related to the measurement of an observable with respect
to the ensemble defined by $A_{\rm tot}$ and $w$. Error bar estimation is an
important aspect of Monte-Carlo simulation. We note that it is
necessary to take into account the covariance between $\left\langle
\hat{O}/w[\{C\}] \right\rangle_{A_{\rm tot},w}$ and $\left\langle
1/w[\{C\}] \right\rangle_{A_{\rm tot},w}$ when determining error
bars. We use the bootstrap method for error analysis.

The form of $w$ is a matter of choice, but it is useful to choose a
form that allows control over the confinement of the new path. To
avoid undesirable long timestep correlations, we use an auxiliary
weighting of the form $w \propto 1/(\alpha + R^{d+1+\eta})$, where $R$
is the distance of the configuration from the impurity, $d$ the
dimensionality of the lattice, $\eta$ is a small value and $\alpha$
stops the weight blowing up on the impurity site. In this way, the
average distance from the impurity is finite, since in the absence of
interaction, $\langle R \rangle \sim \int_{0}^{\infty} w[R] R^{d-1}
\,dR$, shows that ``free'' particles are localized, but not bound too
strongly.

In the path integral QMC for a polaron in an impurity, it is necessary
 to make update operations involving two kinks to ensure that the end
 configurations of the path remain periodic in imaginary time, since
 the translational symmetry has been broken. A binary update satisfying
 the imaginary-time boundary conditions involves adding or removing a
 kink-antikink pair (an antikink to kink $\lvec$ is a kink with
 direction $-\lvec$). Update probability is determined following a
 similar argument to that in Ref. \cite{hague2007b} with a small
 modification; consider two path configurations, $\{C\}$ and $\{D\}$,
 where configuration $\{D\}$ has one more ${\bf l}$ kink at time
 $\tau_1$ and one more $-{\bf l}$ antikink at time $\tau_2$ than
 $\{C\}$.  The balance equation is
$W[\{C\}] Q_A[\{C\}] P[\{C\} \rightarrow \{D\}] = W[\{D\}] Q_R[\{D\}]
P[\{D\} \rightarrow \{C\}] $.
With relative contribution of configurations $\{C\}$ and $\{D\}$
modified by the auxiliary weighting, $W[\{D\}]/W[\{C\}] = (t \Delta
\tau)^2 e^{A[\{D\}]-A[\{C\}]}w[\{D\}]/w[\{C\}]$. 

We modify the equal weighting scheme in Ref. \onlinecite{hague2007b}
for the probability of choosing kinks with a particular kink time when
adding or removing the second kink, allowing weighted kink insertion
with the anti-kink time chosen with probability $p(\tau-\tau')$ where
$\int_{0}^{\beta} p(\tau) d\tau = 1$. In this way, the kinks and antikinks
in the pair can be chosen at similar $\tau$, and the update acceptance is
improved. This choice leads to update probabilities for insertion of a
direction $\lvec$ kink at $\tau$ and a direction $-\lvec$ kink at
$\tau'$,
\begin{equation}
P[C\rightarrow D]
= \min \left\{ 1 \; ; \;
\frac{w[D] t_{\lvec} t_{-\lvec} \beta e^{A[D]-A[C]}} {w[C] N_{{\bf l}}[D] \sum_{i=1}^{N_{-{\bf l}}[D]} p(\tau,\tau_i)}\right\}
\label{eqn:update}
\end{equation}

In order to sample the configuration space faster, we also examined
path updates which shift the whole path through a certain distance,
but the efficiency of the algorithm was not improved. The auxiliary
weighting approach may be used for systems with several particles,
with the auxiliary weight depending on either the absolute or relative
positions of the paths.

In the weighted ensemble, the ground state polaron energy, number of
phonons and average distance from the impurity are respectively
computed via,
\begin{equation}
E =
-\lim_{\beta\rightarrow\infty} \left[\frac{1}{\beta}\left\langle \frac{N}{w} +  \frac{1}{w} \frac{\partial
A}{\partial\beta} \right\rangle_{A_{\rm tot},w} \right]/ \left\langle \frac{1}{w}\right\rangle_{A_{\rm tot},w}
\end{equation}
\begin{equation}
N_{\mathrm{ph}} = - \lim_{\beta\rightarrow\infty}\frac{1}{\bar{\beta}}
\left\langle \frac{1}{w}\left.
\frac{\partial A}{\partial \bar{\omega}}\right|_{\lambda\bar{\omega}}\right\rangle_{A_{\rm tot},w} / \left\langle \frac{1}{w}\right\rangle_{A_{\rm tot},w}\: ,
\label{eq:ten}
\end{equation}
\begin{equation}
R_{{\rm imp}}=\left\langle\frac{1}{w}\sqrt{\frac{1}{\beta}\int_{0}^{\beta}\mathbf{r}^{2}(\tau)
d\tau}\right\rangle_{A_{\rm tot},w}/\left\langle\frac{1}{w}\right\rangle_{A_{\rm tot},w}
\end{equation}
where $N$ is the total number of kinks in the path.

Applying the Lang-Firsov canonical transformation to the Hamiltonian
and assuming large phonon frequency, an approximate form can be
derived,
%
%\begin{equation}
$\tilde{H} = -\tilde{t}\sum_{\langle ij \rangle}c^{\dagger}_{i}c_{j} + \sum_i\Delta_i c^{\dagger}_{i}c_{i}$,
%\end{equation}
%
with $\tilde{t} = t \exp(-z t \lambda/\omega)$. It is possible to
compute an analytic value of the impurity potential, $\Delta$ at which
binding occurs. For polarons on a 3D lattice under the influence of an
impurity, $\Delta_{C} = 3.958 \tilde{t} = 3.958 m_{0} / m^{*}$. When
there is no electron-phonon coupling, the relation $\Delta_{C} = 3.958
t$ is exact. While the approximation for $\tilde{t}$ is not expected
to hold for low phonon frequency, use of the exact value of the
inverse mass computed from CTQMC in the absence of the impurity is
expected to lead to a qualitatively correct value for $\Delta_{C}$. We
will return to this point later in the article. An exact value for the
energy of the instantaneous problem is given by the equation, $1 =
|\Delta|\int\int\int^{\pi}_{-\pi}d^{3}{\bf q}/(2\pi)^3
1/[|E|-2\sum_{i=1}^{d}\cos(q_i)]$.

\begin{figure}
\includegraphics[height=71.5mm,angle=270]{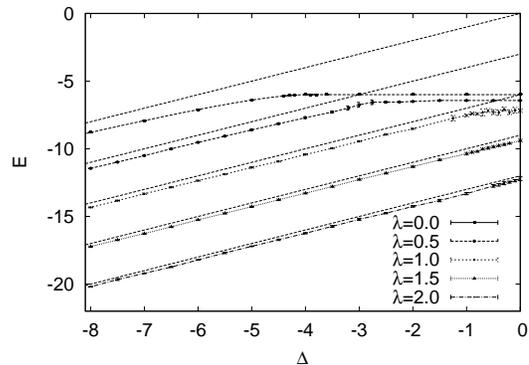}
\caption{Total energy $E$ vs $\Delta$ for various $\lambda$ at
$\bar{\omega} = 1$. In all figures, error bars are plotted to 3
standard errors. Where error bars are not visible, error bars are
smaller than the points. The transition can be seen as a sudden change
in the gradient. The flat gradient corresponds to the energy of the
unbound polaron. Straight dotted lines indicate the strong impurity
asymptote $-zt\lambda + \Delta$}
\label{fig:energy}
\end{figure}

In order to determine the circumstances under which the polaron is
localized by the impurity, the total energy must be determined.
Fig. \ref{fig:energy} shows the total energy $E$ vs $\Delta$ for
various $\lambda$ at $\bar{\omega} = 1$. The transition can be seen as a
sudden change in the gradient. The flat gradient corresponds to the
energy of the unbound polaron. The strong binding asymptotes show how
the small polaron with large $\lambda$ almost immediately binds
strongly as an impurity potential in introduced. We show the exact
energy as a line beneath the $\lambda=0$ points, but for the other
cases, lines are a guide to the eye.

\begin{figure}
\includegraphics[height=71.5mm,angle=270]{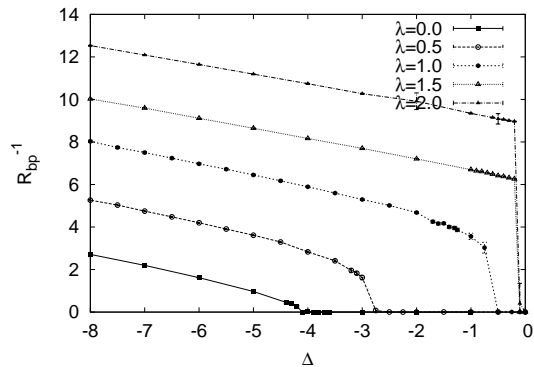}
\caption{$R_{\rm imp}^{-1}$ vs $\Delta$ for various $\lambda$ at $\bar{\omega} =
1$.}
\label{fig:inverseradius}
\end{figure}

The measure of the inverse radius gives another criterion for the
critical binding potential. Fig. \ref{fig:inverseradius} shows $R_{\rm
imp}^{-1}$ vs $\Delta$ for various $\lambda$ at $\bar{\omega} =
1$. The inverse radius vanishes continuously on binding. As the
electron-phonon interaction increases in strength, the polaron binds
at monotonically smaller values of the impurity potential, until
around $\lambda=1$, where tiny impurities are capable of binding the
polaron. This is because the kinetic energy of the polaron decreases
rapidly on increasing $\lambda$. Binding occurs when the depth of
impurity potential is approximately the kinetic energy of the free
polaron.

\begin{figure}
\includegraphics[height=71.5mm,angle=270]{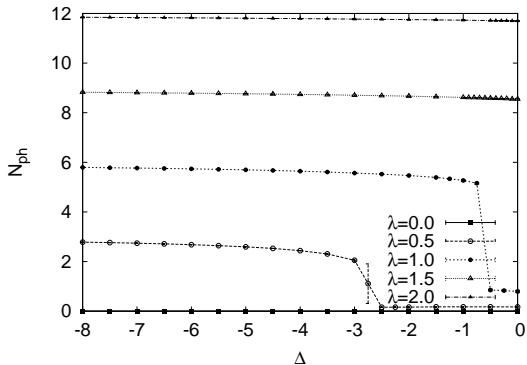}
\caption{$N_{ph}$ vs $\Delta$ for various $\lambda$ at $\bar{\omega} =
1$. For small $\lambda$, there is a sudden increase in $N_{ph}$ as the
large polaron is trapped by the impurity, leading to small polaron
behavior.}
\label{fig:nphonons}
\end{figure}

Small (strong-coupling) polarons have much larger numbers of phonons
associated with the electron than large (weak-coupling) polarons. It
is therefore of interest to see if the reduction in size associated
with localization is also associated with a similar
phenomenon. Fig. \ref{fig:nphonons} shows the variation of $N_{ph}$
with $\Delta$ at a number of different $\lambda$ at $\omega/t =
1$. For small $\lambda$, there is a sudden increase in the number of
phonons as the large polaron is trapped by the impurity,  to
small polaron behavior. The localization of the polaron wavefunction
increases the possibility for interaction between the electron and the
local ion mode through a local coupling. Such an effect is expected to
be less pronounced for long range electron-phonon coupling.

\begin{figure}
\includegraphics[height=71.5mm,angle=270]{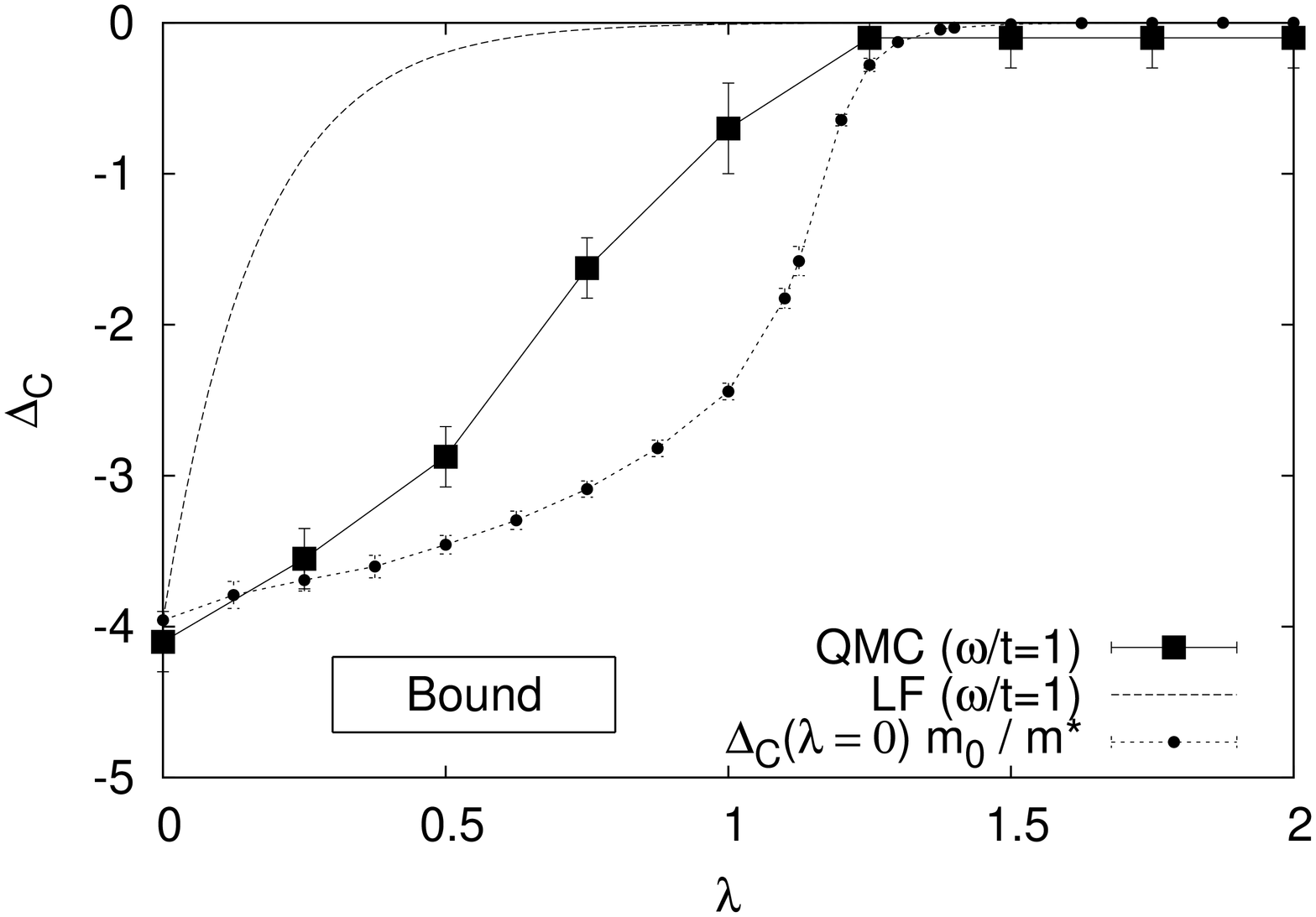}
\includegraphics[height=71.5mm,angle=270]{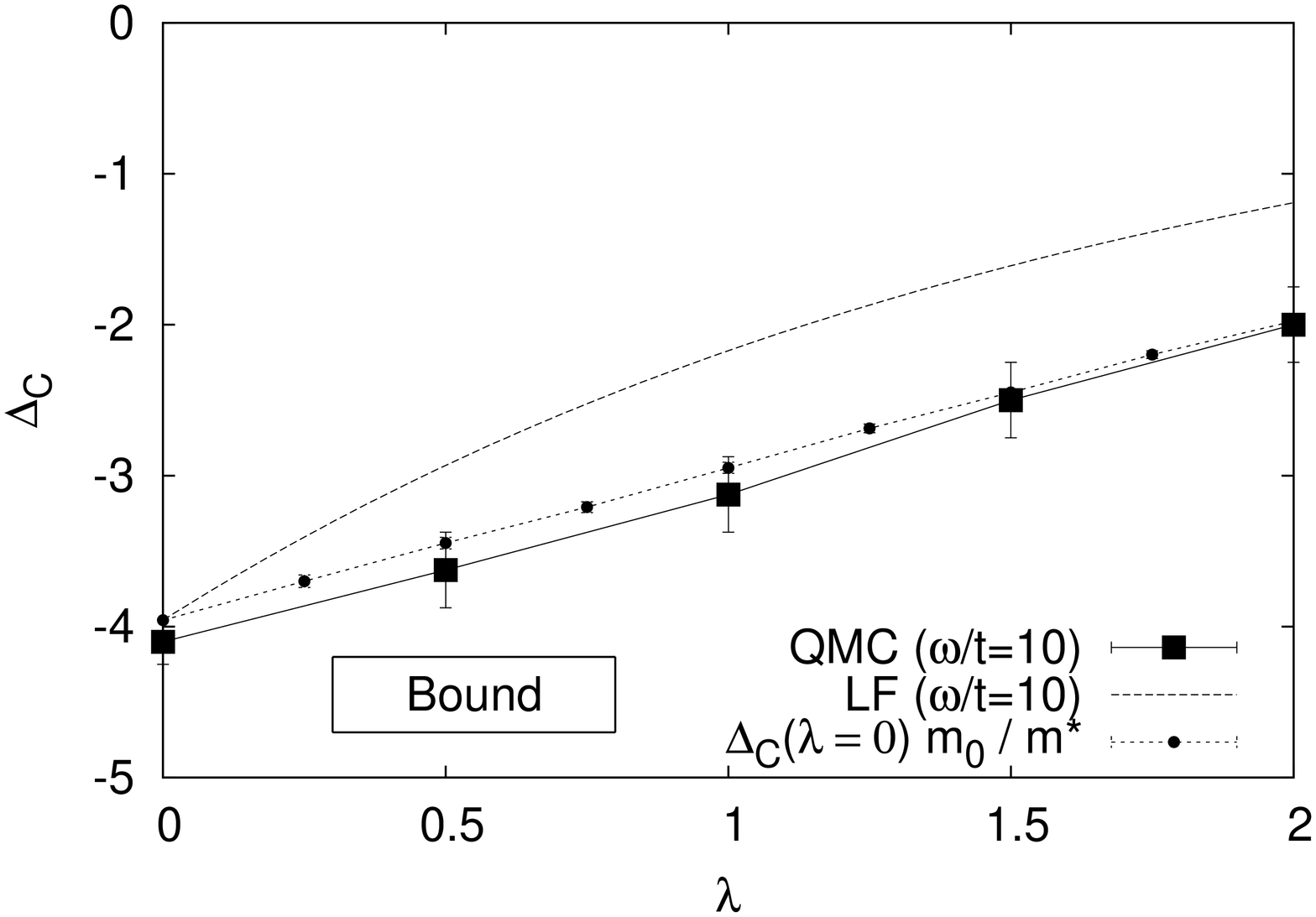}
\caption{Binding diagram in $\lambda,\Delta$ space. Also shown are the
approximate values predicted by the Lang-Firsov transformation with
zero excited phonons $\Delta_C(\lambda) =
\Delta_C(0)\exp(-z\lambda /\bar{\omega})$, from the exact mass in the
case with translational invariance (measured using QMC) $\Delta_C(\lambda) =
\Delta_C(0)m_0/m^{*}$ and the value measured using
Monte-Carlo.}
\label{fig:bindingdiagram}
\end{figure}

Finally, we show the binding diagram in $\lambda,\Delta$ space in
Fig. \ref{fig:bindingdiagram}. Also shown are the approximate values
predicted by the Lang--Firsov transformation with zero excited phonons
$\Delta_C(\lambda) = \Delta_C(0)\exp(-z\lambda/\bar{\omega})$, from the
exact mass in the translationally-invariant case (measured using QMC)
$\Delta_C(\lambda) = \Delta_C(0)m_0/m^{*}$ and the value measured from
the current Monte-Carlo code. For $\bar{\omega}=1$, we
examined the energy for $\bar{\beta} = 112$ close to binding. Below
$\bar{\beta} = 56$, error bars dominated temperature corrections, and
led to small differences between analytics and numerics when
$\bar{\omega}=10$. At low phonon frequencies, the Holstein polaron
binds almost immediately at $\lambda\gtrsim 1$. This leads to the
expectation that polarons with local interaction are almost completely
localized in materials with strong electron-phonon coupling. Mobile
behavior could re-emerge if the electron density is similar to the
density of impurities and a strong Coulomb repulsion ($U\gtrsim
4\lambda$) is also available so that polarons become repulsive rather
than attractive impurities when pinned. Longer range interactions lead
to more mobile polarons, and therefore less likely to be pinned (the
role of mass is shown by the similarity between the binding diagram
from QMC, and the approximate one determined from the mass of the
unbound polaron). Such long range electron-phonon interactions are
difficult to justify in 3D materials, and we expect that polarons
strongly coupled with low-frequency phonons are strongly localized in such
materials.

In summary, we have developed an algorithm for studying the trapping
of polarons by impurities. Analysis of the total energy showed that
for electron-phonon coupling $\lambda>1$, polarons are strongly bound
to the impurity. For small $\lambda$, there is a more gradual binding,
coupled with a sudden increase in the number of phonons present in the
polaron. We computed the binding diagram, showing that critical
impurity strength changes monotonically with $\lambda$ for
intermediate-coupling, in contrast to the conclusion for continuum
(weak-coupling) polarons, and differs significantly from the
strong-coupling approximation.

This work was supported by EPSRC (UK) (grant EP/C518365/1). We thank
Andrei Mishchenko and John Samson for valuable discussions.

%\bibliography{disordered_polaron_prlREvised}

\end{document}